\documentclass[12pt]{emulateapj}
\usepackage{amsmath,amssymb}
\usepackage[usenames,dvipsnames]{xcolor}
\usepackage{graphicx}
\usepackage{wasysym}
\newcommand{\dd} {\mathrm{d}}
\newcommand{\p}{\text{p}}
\slugcomment{\today}

\begin{document}

\title{On the Generation of Compressible Mirror-mode Fluctuations in the Inner Heliosheath}

\author{Horst Fichtner$^{1,2}$, Jens Kleimann$^1$, Peter H.\ Yoon$^{3,4,5}$, Klaus Scherer$^{1,2}$, Sean Oughton$^6$, N. Eugene Engelbrecht$^{7,8}$}
\affiliation{$^1$Institut f\"ur Theoretische Physik IV, Ruhr-Universit\"at Bochum, 44780 Bochum, Germany}
\affiliation{$^2$Research Department Plasmas with Complex Interactions, Ruhr-Universit\"at Bochum, Germany}
\affiliation{$^3$Institute for Physical Science and Technology, University of Maryland, 
College Park, USA}
\affiliation{$^4$School of Space Research, Kyung Hee University, Yongin, South Korea}
\affiliation{$^5$Korea Astronomy and Space Science Institute, Daejeon 34055, Korea}
\affiliation{$^6$Department of Mathematics and Statistics, University of Waikato, Hamilton 3240, New Zealand}
\affiliation{$^7$Centre for Space Research, North-West University, Potchefstroom, 2522, South Africa}
\affiliation{$^8$National Institute for Theoretical Physics (NITheP), Gauteng, South Africa}

\begin{abstract}
  Measurements made with the Voyager~1 spacecraft indicate that significant levels of compressive fluctuations exist in the inner heliosheath.
  Some studies have already been performed with respect to the mirror-mode instability in the downstream region close to the solar wind termination shock, and here we extend the investigation to the whole inner heliosheath.
  We employ quasilinear theory and results from a global magnetohydrodynamic model of the heliosphere to compute the time evolution of both the temperature anisotropy and the energy density of the corresponding magnetic fluctuations, and we demonstrate their likely presence in the inner heliosheath.
  Furthermore, we compute the associated, locally generated density fluctuations.
  The results can serve as inputs for future models of the transport of compressible turbulence in the inner heliosheath. 
\end{abstract} 


\section{Introduction}
Voyager spacecraft observations \citep[e.g.,][]{Richardson-Burlaga-2013} indicate that in the inner heliosheath (IHS) the small scales display considerable variability and have a significant compressible component.
The existence of the latter has been established via analyses of Voyager~1 and~2 data in a series of papers by Burlaga and co-workers \citep{Burlaga-etal-2006, Burlaga-etal-2014, Burlaga-Ness-2012b, Burlaga-Ness-2012a} and, more recently, by \citet{Fraternale-etal-2019}. 
This compressibility of the fluctuations in the IHS is in contrast to the situation in the solar wind before it encounters its termination shock, where the fluctuations are predominantly incompressible \citep[e.g.,][]{Tu-Marsch-1994, Roberts-etal-2018}.
Attempts to model compressive fluctuations and their transport in the IHS are at an early stage.
It seems likely that they are generated at and/or near the termination shock \citep[e.g.,][]{Adhikari-etal-2016, Zank-etal-2018}, and also within the heliosheath \citep[e.g.,][]{Fahr-Siewert-2007, Liu-etal-2007, Liu-etal-2010, Genot-2008, Genot-2009, Tsurutani-etal-2011a, Tsurutani-etal-2011b}.

Even for the upstream heliosphere, i.e., for the region enclosed by the solar wind termination shock, a rigorous one-component theory of the transport of compressive fluctuations has been developed only relatively recently \citep{Hunana-etal-2008, Hunana-Zank-2010, Zank-etal-2012b, Zank-etal-2013b, Zank-etal-2017} to explain observations reviewed by \citet{Bruno-Carbone-2016}.
The only work that systematically and quantitatively attempted to model the turbulence transport within the IHS has been presented by \citet{Usmanov-etal-2016}.
However, providing the first ``rigorous'' approach to the solution of this problem, these authors justifiably made the simplest possible but already involved first step by applying the well-established one-component model for incompressible turbulence in this region.
While being well aware that not considering compressive fluctuations is a serious limitation, \citet{Usmanov-etal-2016} demonstrated the principal feasibility of such an extension of previous transport modelling and provided a valuable reference case for forthcoming simulations.

Future models of turbulence transport in the IHS, however, will need to include the compressive fluctuations. 
Although transmission of fluctuations from the upstream (supersonic) solar wind through the termination shock into the downstream flow can account for some of the compressive fluctuations, it will also be necessary to include their
generation at the shock \citep{Zank-etal-2010} and in the IHS \citep{Liu-etal-2007, Liu-etal-2010}. 
In particular, the latter aspect, i.e., the production of compressible turbulence within the IHS has, to the best of our knowledge, not yet been studied systematically for the whole IHS.
With the present paper, we begin such a study, the results of which can subsequently be used in models of the transport of compressive fluctuations in this region of the heliosphere.
In order to tackle this, we apply quasilinear theory to follow the evolution of the mirror-mode instability using initial values that are obtained from a simulation of a three-dimensional (3D) model of the large-scale heliosphere. 

The further structure of the paper is as follows.
In Section~\ref{sec:2} we briefly summarize how the temperature anisotropy, magnetic field fluctutations, and density fluctuations are related for the mirror-mode instability.
In Section~\ref{sec:3} we first outline the quasilinear theory employed to follow the time evolution of the temperature anisotropy and the energy density of the magnetic fluctuations, and then describe the simulation model used to determine the plasma and magnetic field structure of the IHS.
In Section~\ref{sec:4} we apply both models to the generation of compressive fluctuations in the IHS, and in Section~\ref{sec:5} we summarize all results and draw some conclusions. \\

\section{Compressive Fluctuations Due to the Mirror-mode Instability}
\label{sec:2}
It is well known that the proton mirror instability generates compressive fluctuations, see, e.g., \citet{Hasegawa-1969}, 
\citet{Qu-etal-2007}, or \citet{Hellinger-etal-2017}.
This instability is a consequence of a temperature anisotropy $A := T_\perp/T_\| > 1$ (with $\|,\perp$ referring to the orientation relative to the local magnetic field direction $\vec{B}/B$), and the resulting fluctuations $\delta n_\p$ in proton number density $n_\p$ are anticorrelated with the associated magnetic fluctuations $\delta B$ via the relation \citep[e.g.,][]{Liu-etal-2007}
\begin{equation}
  \label{deltan}
  \frac{\delta n_\p}{n_\p} = -(A -1)\,\frac{\delta B}{B} .
\end{equation}
While \citet{Fahr-Siewert-2007}, \citet{Liu-etal-2007}, \citet{Genot-2008}, and \citet{Liu-etal-2010} have predicted the mirror-mode fluctuations mainly in the downstream vicinity of the solar wind termination shock, \citet{Tsurutani-etal-2011a} have not only experimentally verified these predictions with Voyager~1 data, but have also pointed out that the injection of pickup ions throughout the heliosheath will lead to a further mirror-mode amplification \citep[see also][]{Tsurutani-etal-2011b}.
This has been further confirmed by \citet{Burlaga-Ness-2011}, who demonstrated that the data of the Voyager~1 spacecraft are consistent with a mirror-mode instability deep into the heliosheath.

Following up on these ideas, we apply in the following the theory of temperature anisotropy-driven kinetic instabilities \citep{Yoon-2017} to the IHS. 
The structure of the latter is obtained from a numerical simulation of the 3D interaction of the solar wind with the local interstellar medium (LISM) using the MHD code \textsc{Cronos} \citep{Kissmann-etal-2018}. 

\section{The model}
\label{sec:3}
In the following two subsections, first, we briefly review the quasilinear theory of the mirror-mode instability as presented in \citet{Yoon-2017} and second, we describe the MHD model used to obtain the structure of the IHS.

\subsection{Quasilinear Theory of the Mirror-mode Instability}
\label{qltmodel}
The quasilinear time evolution of the temperature parallel and perpendicular to the magnetic field is given by \citep{Yoon-2017}
\begin{eqnarray}
  \frac{\dd T_{a \|}}{\dd t} &=& \frac{2\pi m_a}{k_\text{B}}\int  v_\|^2 \frac{\partial F_a}{\partial t}\, v_\perp \dd v_\perp \dd v_\| \\
  &=& -\frac{4\pi m_a}{k_\text{B}} \int v_\| \left(D_{\|\perp} \frac{\partial F_a}{\partial v_\perp}
  + D_{\|\|}    \frac{\partial F_a}{\partial v_\|}\right)\, v_\perp \dd v_\perp \dd v_\|, \nonumber\\
  & & \nonumber\\
  \frac{\dd T_{a \perp}}{\dd t} &=& \frac{\pi m_a}{k_\text{B}}\int  v_\perp^2 \frac{\partial F_a}{\partial t}\, v_\perp \dd v_\perp \dd v_\| \\
  &=& -\frac{2\pi m_a}{k_\text{B}} \int v_\perp \left(D_{\perp\perp} \frac{\partial F_a}{\partial v_\perp}
  + D_{\perp\|}    \frac{\partial F_a}{\partial v_\|}\right)\, v_\perp \dd v_\perp \dd v_\| , \nonumber
\end{eqnarray}
where integration by parts has been used, $k_\text{B}$ is the Boltzmann constant, and $m_a$ is the mass of particles of species $a$ with the velocity distribution function $F_a$ that is normalized to unity and obeys the Fokker--Planck type equation
\begin{eqnarray}
  \frac{\partial F_a}{\partial t} &=& \frac{\partial}{\partial v_\|}
  \left(D_{\|\perp} \frac{\partial F_a}{\partial v_\perp}
  + D_{\|\|}    \frac{\partial F_a}{\partial v_\|}\right)\nonumber\\
  &&+ \frac{1}{v_\perp}\frac{\partial}{\partial v_\perp}
  \left[v_\perp \left(D_{\perp\perp} \frac{\partial F_a}{\partial v_\perp}
    + D_{\perp\|} \frac{\partial F_a}{\partial v_\|}\right)\right] ,
\end{eqnarray}
wherein the coefficients have the form 
\begin{eqnarray}
  D_{\alpha\beta} &=& i\frac{q_a^2}{m_a^2} 
  \int \sum\limits_{n=-\infty}^{\infty}\frac{1}{\omega-n\Omega_a-k_\|v_\|} \nonumber\\
  &&\times \left(\left\langle\left\vert\frac{nJ_n(\kappa_a)}{\kappa_a}\,\delta E_{\vec{k},x} 
  + iJ^{\,\prime}_n(\kappa_a) \, \delta E_{\vec{k},y}\right\vert^2\right\rangle 
  \Delta_\alpha^*\Delta_\beta\right. \nonumber\\
  & & \hspace*{1.0cm} \left. \vphantom{\sum\limits_{n=-\infty}^{\infty}} 
  + J_n^2(\kappa_a) \langle\delta E_{\vec{k},z}^2\rangle \Lambda_\alpha^*\Lambda_\beta\right)\, \dd^3 k ,
  \label{coeffs}
\end{eqnarray}
with the abbreviations
\begin{eqnarray}
  \kappa_a := \frac{k_\perp v_\perp}{\Omega_a} , \quad
  \Delta_\| := \frac{k_\| v_\perp}{\omega}    , \quad 
  \Delta_\perp := 1 - \frac{k_\| v_\|}{\omega} ,
  \label{abbrev1}
\end{eqnarray}
and
\begin{eqnarray}
  \Lambda_\| := 1 - \frac{n\Omega_a}{\omega} ,
  \quad 
  \Lambda_\perp := \frac{n\Omega_a}{\omega}\frac{v_\|}{v_\perp} .
  \label{abbrev2}
\end{eqnarray}
In Equations~(\ref{coeffs}), (\ref{abbrev1}), and (\ref{abbrev2}) $q_a$ is the charge, $\Omega_a$ the gyro frequency of particles of species $a$, $\vec{k}$ and $\omega$ are the wave vector and frequency of the fluctuating electric field $\delta \vec{E}$ with $z$-component along $\vec{B}$, and $J_n(x)$ denotes the Bessel function of the first kind and order $n$. 

The evaluation of the integrals for the temperature evolution is carried out as follows:
\begin{itemize}
\item[(i)] We make the simplifiying assumption that the velocity distributions are bi-Maxwellians, i.e.,
  \begin{equation}
    F_a(v_\|,v_\perp) = \frac{1}{\pi^{3/2}\alpha_\| \, \alpha_\perp^2}                    \exp\left(-\frac{v_\|^2}{\alpha_\|^2}-\frac{v_\perp^2}{\alpha_\perp^2}\right) ,
  \end{equation}
  with the thermal speeds $\alpha_{\|,\perp} := \sqrt{2 k_\text{B} T_{a \|,a \perp}/m_a}$.
\item[(ii)] For the low-frequency mirror mode we may consider only the $y$-component of the electric field and the Landau resonance $n=0$, i.e., we neglect all cyclotron harmonics. 
\item[(iii)] We express the linear dispersion relation for the mirror-mode instability in an electron-proton plasma \mbox{($a=\p$)} in the form 
  \begin{equation}
    \frac{c^2k^2}{\omega_\text{pp}^2} = -2\lambda(\Lambda_0-\Lambda_1)\left(1+\frac{T_{\p\perp}}{T_{\p\|}}\frac{Z^{\,\prime}(\xi)}{2}\right) ,
  \end{equation}
  where $c$ is the speed of light, $\omega_\text{pp}$ the  proton  plasma frequency, $\lambda := \kappa_\p^2/2$, and $T_{\p\|,\p\perp}$ the temperature components of the protons, $\Lambda_{n}(\lambda) := I_n(\lambda) \exp(-\lambda)$ with $I_n(x)$ denoting the modified Bessel function of the first kind and order $n$, and $Z^{\,\prime}(\xi) := (2/\sqrt{\pi}) \int_{-\infty}^\infty y\exp(-y^2)/(\xi-y) \, \dd y$ with the variable $\xi := \omega/(k_\| \alpha_{\|})$.
\end{itemize}

Employing (i) to (iii) as in \citet{Yoon-2017} and using $\langle\delta B_\text{mm}^2\rangle_{\vec{k}} = \vert c^2k^2/\omega^2\vert \langle \delta E^2 \rangle_{\vec{k}}$ results in
\begin{eqnarray}
  \frac{\dd(2\mu_0 n_\p k_\text{B} T_{\p\|}/B^2)}{\dd (t\;\Omega_\p)} &=& \frac{2e^2}{m_\p} \frac{2\mu_0 n_\p}{B^2} \\ 
  && \hspace*{-2cm} \times \int \frac{\gamma_{\vec{k}}}{\Omega_\p} \frac{\langle\delta B_\text{mm}^2\rangle_{\vec{k}}}{c^2k^2}
  \left(2\lambda (\Lambda_0-\Lambda_1) + \frac{c^2k^2}{\omega_\text{pp}^2}\right) \dd^3 k , \nonumber\\
  && \nonumber \\
  \frac{\dd(2\mu_0 n_\p k_\text{B} T_{\p\perp}/B^2)}{\dd (t\;\Omega_\p)} &=& -\frac{2e^2}{m_\p} \frac{2\mu_0 n_\p}{B^2} \\ 
  && \hspace*{-2cm} \times \int \frac{\gamma_{\vec{k}}}{\Omega_\p} \frac{\langle\delta B_\text{mm}^2\rangle_{\vec{k}}}{c^2k^2}
  \left(\lambda (\Lambda_0-\Lambda_1) + \frac{c^2k^2}{\omega_\text{pp}^2}\right) \dd^3 k ,
  \nonumber
\end{eqnarray}
with the elementary charge $e$, the permeability of vacuum $\mu_0$, and $\gamma_{\vec{k}}$ being the imaginary part of $\omega$.
With the definitions of the plasma betas, a normalized wave vector, the normalized fluctuation energy density, and normalized time
\begin{equation}
  \begin{array}{rlrl}
    \beta_{\|,\perp} &:= \displaystyle \frac{2\mu_0 n_\p k_\text{B} T_{\p\|,\p\perp}}{B^2} , &
    \vec{q} &:= \displaystyle \frac{c\, \vec{k}}{\omega_\text{pp}} \\
        {\cal W}_\text{mm} &:= \displaystyle \frac{\langle (\delta B_\text{mm})^2 \rangle_{\vec{k}}}{B^2} , & \tau &:= t\,\Omega_\p ,
        \label{enedens}
  \end{array}
\end{equation}
one obtains the dimensionless forms 
\begin{eqnarray}
  \frac{\dd\beta_{\|}}{\dd\tau} &=& 4 \int \frac{\omega}{\Omega_\p} \left(\frac{2\lambda(\Lambda_0-\Lambda_1)}{q^2}+1 \right)\,{\cal W}_\text{mm}\, \dd^3q , \\
  \frac{\dd\beta_{\perp}}{\dd\tau} &=& -4 \int \frac{\omega}{\Omega_\p} \left(\frac{\lambda(\Lambda_0-\Lambda_1)}{q^2}+1 \right)\,{\cal W}_\text{mm}\, \dd^3q .
\end{eqnarray}
Note that $\lambda = q_\perp^2\beta_\perp/2$ and that, as a consistency check, energy conservation is fulfilled as the relation 
\begin{eqnarray}
  \frac{\dd}{\dd\tau}\left(\frac{\beta_\|}{2} + \beta_\perp\right)
  = -2 \int \frac{\omega}{\Omega_\p}\, {\cal W}_\text{mm}\, \dd^3q 
\end{eqnarray}
holds.

\begin{figure}[h]
  \includegraphics[width=1.0\columnwidth]{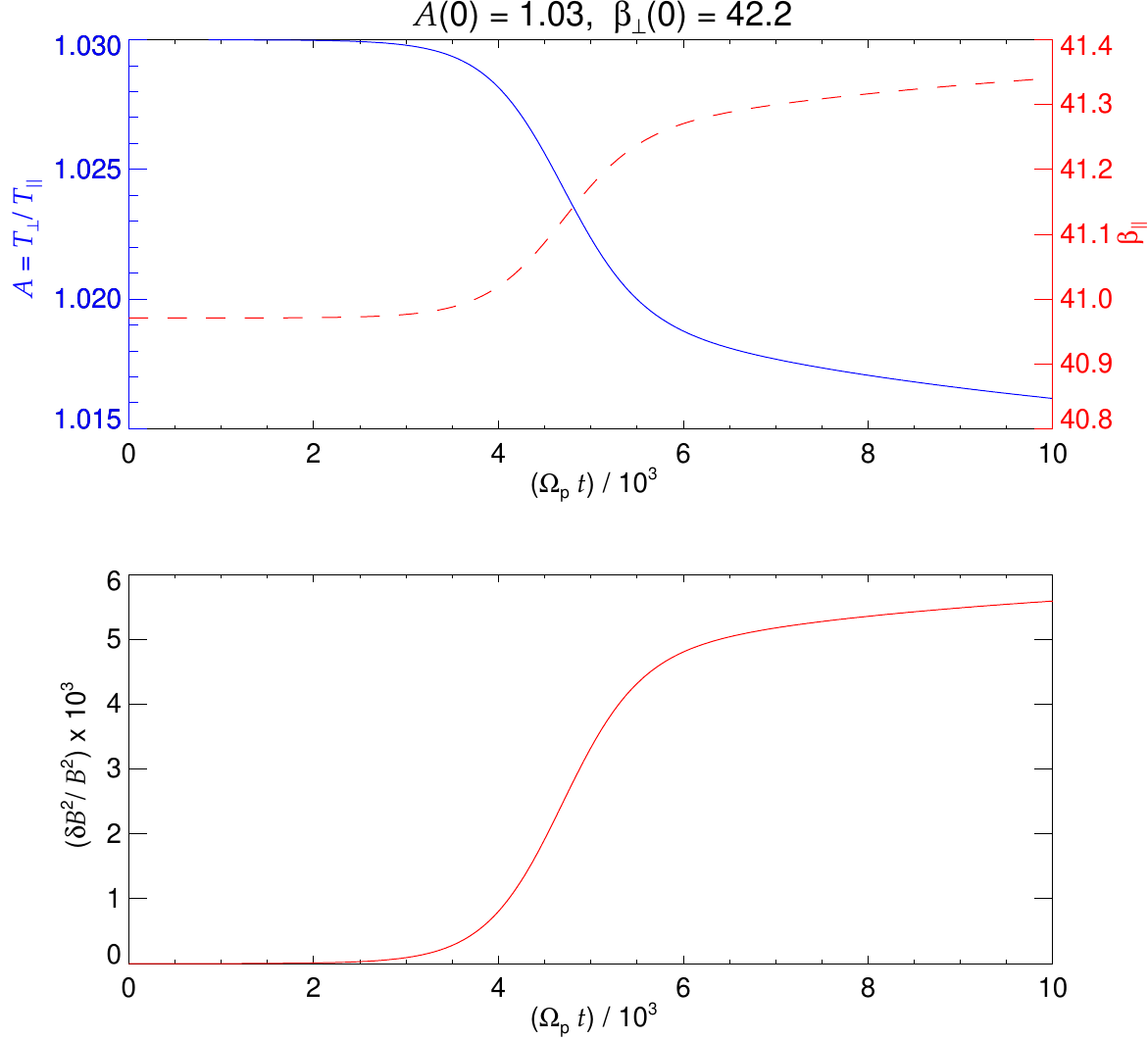}
  \caption{The quasilinear time evolution of the mirror-mode instability: the temperature anisotropy (blue) and parallel plasma beta (red, upper panel), and the normalized energy density of the fluctuations (lower panel).\\
  }
  \label{fig:example}
\end{figure}

As an illustrative example, we show the quasilinear evolution of the perpendicular temperature anisotropy and the normalized fluctuation energy for one set of initial values in Figue~\ref{fig:example}.
These observationally guided initial values $A(0) = 1.03$ and $\beta_\perp(0) = 42.2$ at $\tau = 0$ are taken from \citet{Liu-etal-2007}.

Evidently, the temperature anisotropy/parallel plasma beta is quasi-saturating at a level somewhat lower/higher than the initial values.
The fluctuation energy density is, after an initially slow increase, quickly increasing to its quasi-saturation level after a few thousand gyroperiods.
Such run times were used for all quasilinear calculations discussed in the following.

\subsection{The MHD Model of the Large-scale Heliosphere and the Structure of the IHS}
\label{sec:mhdmodel}
The \textsc{Cronos} code \citep{Kissmann-etal-2018} was used to integrate the equations of ideal one-fluid MHD augmented by additional equations for three quantities characterizing the fluctuations.
These are twice the total fluctuation energy per unit mass \mbox{$Z^2 := \left< (\delta u)^2 \right> + \left< (\delta B)^2/\rho \right>$}, the normalized cross-helicity \mbox{$\sigma_c := 2 \big< \delta\vec{u} \cdot \delta\vec{B}/\rho \big>/Z^2$}, and a characteristic length scale $\lambda$, as described in and used by \citet{Wiengarten-etal-2015} and others. 
Because the equations for the turbulence quantities and the large-scale flow are 
(i) coupled self-consistently and
(ii) valid in regions where the flow is \emph{sub}-Alfv\'enic (as in the IHS), the model provides a suitable first approximation for describing the evolution of the plasma as it moves across the termination shock and into the IHS.
Note that, as in \citet{Wiengarten-etal-2015}, the simulated $\delta B$ is an approximation for all the incompressible magnetic fluctuations (e.g., from quasi-2D and Alfv\'en wave-like components), but not those attributed to the mirror-mode $\delta B_\text{mm}$, which are accounted for separately here. 
So, one can distinguish the latter's (normalized) energy density ${\cal W}_\text{mm}$ from that of these incompressible MHD fluctuations defined as
\begin{equation}
  {\cal W}_\text{MHD}:=\frac{\left<(\delta B)^2\right>}{B^2}
  = \frac{2}{3} \frac{\rho \, Z^2}{B^2} ,
  \label{eq:W-mhd}
\end{equation}
when assuming that the magnetic and velocity fluctuations are related in magnitude by $\left<(\delta B)^2\right> = 2\rho \left<(\delta u)^2\right>$ (i.e., an Alfv\'en ratio of 1/2).

We used a 3D spherical grid establishing $N_r \times N_{\vartheta} \times N_{\varphi} = 250 \times 60 \times 90$ cells with uniform extensions $[\Delta r,\Delta\vartheta,\Delta\varphi]$ in $[r,\vartheta,\varphi]$ coordinate space, extending over the radial interval $r \in [80,1000]$\,au and full $(\vartheta, \varphi) \in [0,\pi] \times [0,2\pi]$ angular coverage.
The inner boundary conditions at 80\,au were interpolated from the converged state of another simulation of the same setting but extending only to 100\,au in radius.
The respective solar wind boundary conditions at $r=1$\,au for number density and temperature for this innermost grid were $n_\text{sw} = 7$\,cm$^{-3}$ and $T_\text{sw} = 73,640$\,K.
We used a bimodal solar wind whose magnitude changes from 740\,km\,s$^{-1}$ at the poles to 320\,km\,s$^{-1}$ at the solar equator, with a smooth transition around $\pm 15^{\circ}$ of latitude.
The solar magnetic field was a bipolar Parker-type spiral field of radial strength 3.15\,nT.
\begin{figure*}[th]
  \begin{center}
    \includegraphics[width=0.98\textwidth]{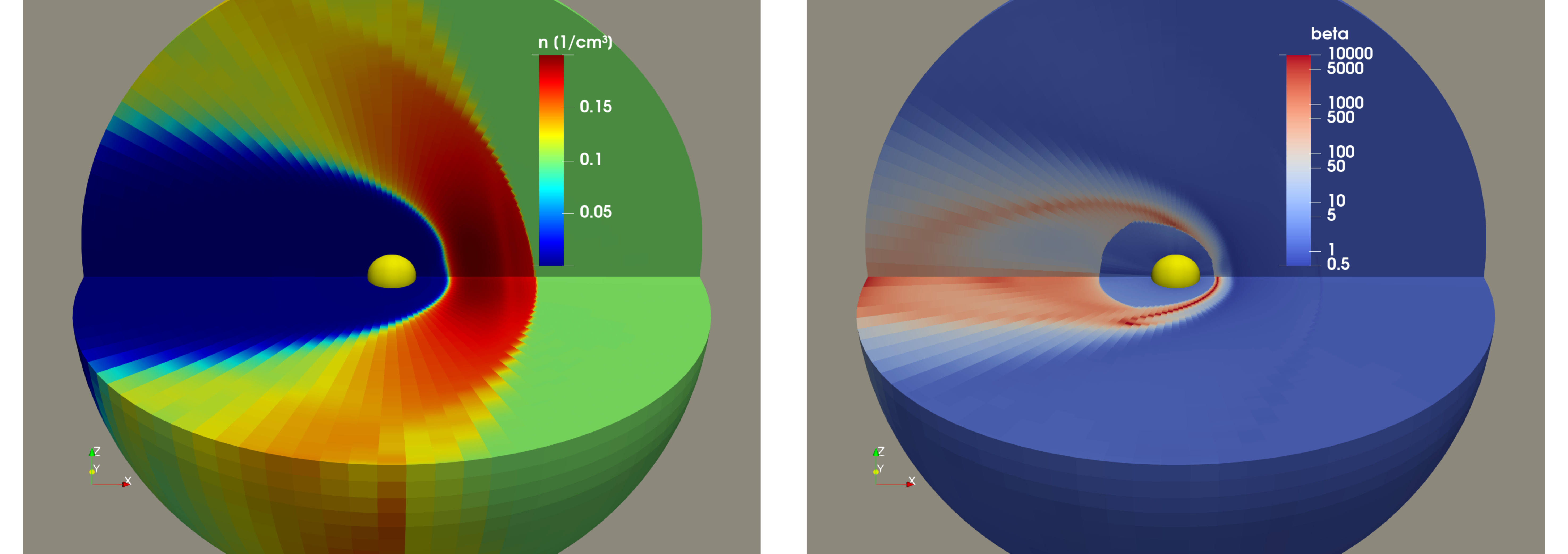}
  \end{center}
  \caption{The simulated model heliosphere visualized with the number density (left panel) and the resulting plasma beta (right panel).
    The heliopause and bow shock are evident in the density plot, while the color code for the beta plot is chosen such that it reveals the termination shock and the internal structure of the IHS.
    The yellow central sphere of 80\,au radius marks the grid's inner boundary in each case. \\}
  \label{fig:heliosphere}
  ~\vspace*{0.1cm}\\
\end{figure*}
The interstellar boundary conditions on the upwind half-space were $n_\text{ism} = 0.1$\,cm$^{-3}$, $T_\text{ism} = 6135$\,K, $v_\text{ism}=26$\,km\,s$^{-1}$, and $B_\text{ism}=0.2$\,nT, with the respective orientations of $\vec{v}_\text{ism}$ and $\vec{B}_\text{ism}$ adopted from \citet{Usmanov-etal-2016}, who used results from \citet{Funsten-etal-2013}.
The turbulence quantities were initialized as in \citet{Wiengarten-etal-2015}.
At the boundary of the downwind half-space, all quantities were extrapolated outward, with radial velocity restricted to non-negative values.

The equations were integrated in time until a quasi-steady state was reached after about 400\,yr of simulated time had elapsed.
The final configuration is illustrated in Figure~\ref{fig:heliosphere}.
\begin{figure}[h!]
  \includegraphics[width=1.0\columnwidth]{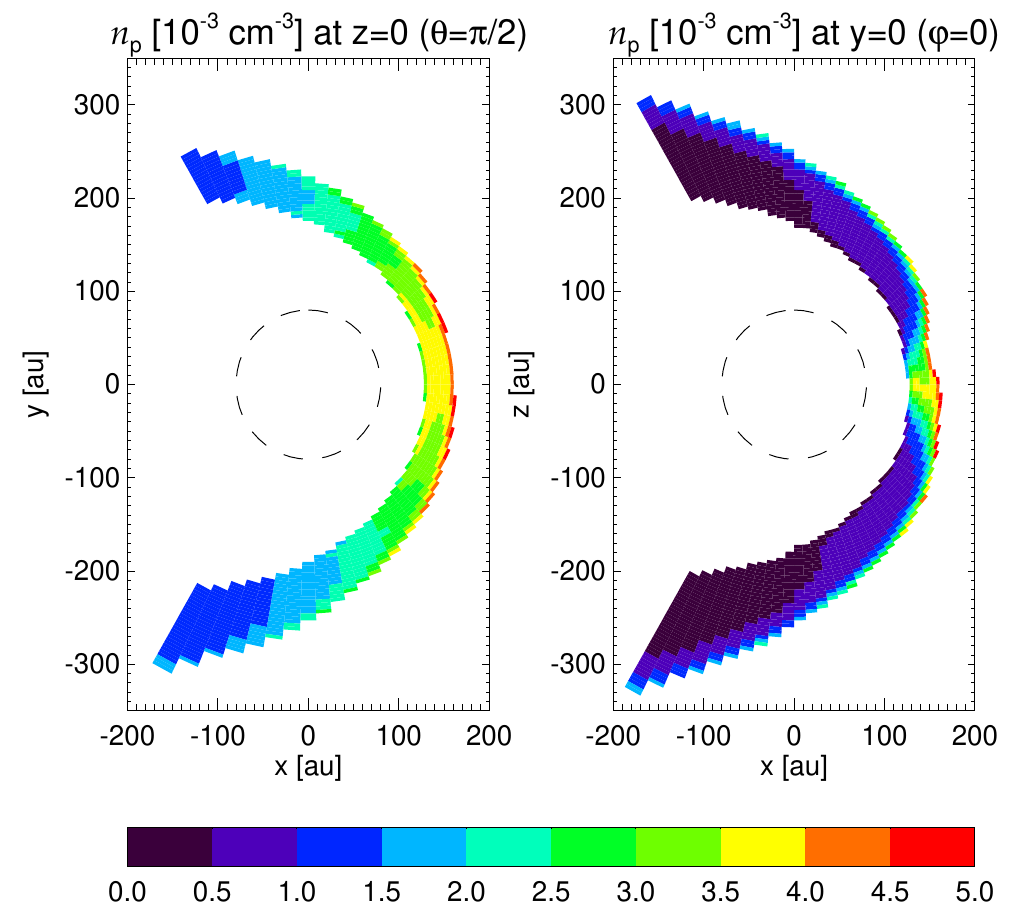}
  \caption{The number density shown in Figure~\ref{fig:heliosphere} limited to the IHS (mainly in the upwind region) in the equatorial plane with $\vartheta=\pi/2$ (left) and the meridional plane with $\varphi=0$ (right).
    The dashed circle indicates the inner boundary of the computational domain at $r = 80$\,au. \\
  } 
  \label{fig:IHS}
\end{figure}
We are focusing in this study on the IHS mainly in the upwind region of the heliosphere.
Therefore, in the following the results are shown exclusively for cells that
(i) have a plasma temperature of at least $10^6$\;K and
(ii) are located outside of a Sun-centered, heliotail-aligned cone of $120^{\circ}$ opening angle.
The locations of cells satisfying these conditions are illustrated in Figure~\ref{fig:IHS}, which displays the number density that was already given in Figure~\ref{fig:heliosphere}.
The asymmetry of the 3D structure, particularly visible in the left panel, is a consequence of the respective undisturbed ISM vectors of velocity and magnetic field meeting at an angle of $\sim\,132^{\circ}$. \\

\begin{figure*}[h!]
  \begin{center}
    \includegraphics[width=1.00\textwidth]{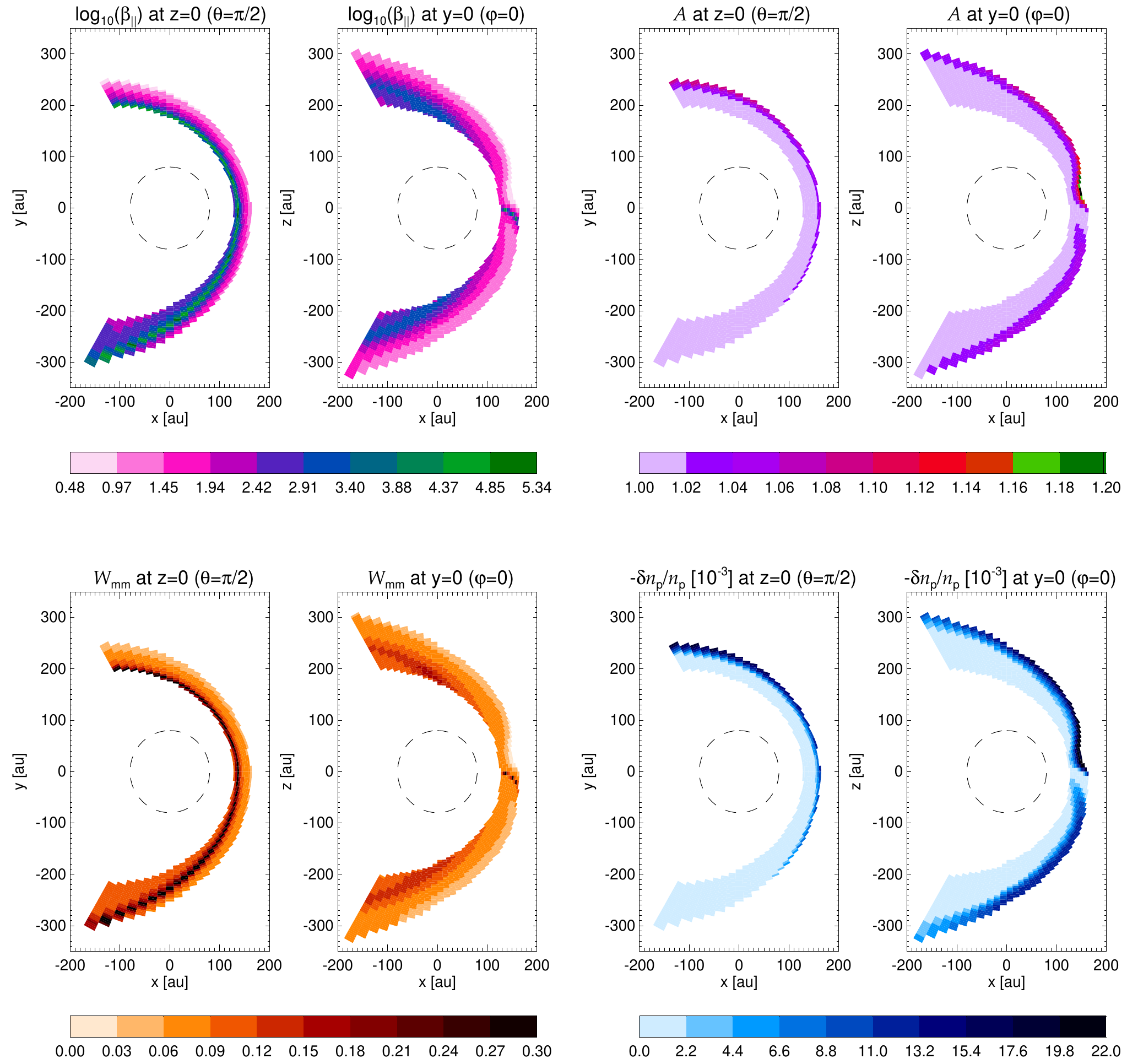}
  \end{center}
  \caption{The quasilinear evolution of the mirror-mode instability in the IHS plotted in the format of Figure~\ref{fig:IHS}, i.e., for each quantity in the equatorial plane ($\vartheta=\pi/2$) and in the meridional plane with $\varphi=0$: the final $\beta_\|$ (upper left panel) the final temperature anisotropy $A$ (upper right panel), the (normalized) energy density ${\cal W}_\text{mm}$ of the magnetic fluctuations according to Equation~(\ref{enedens}) (lower left panel), and the corresponding (normalized) amplitude of the corresponding fluctuations in number density (lower right panel).
    The thin equatorial streak of very high beta values (green stripe in the upper left panel) occurs in an interface region slightly beyond the termination shock characterized by very small magnetic field, which is likely to be an artifact of finite numerical resolution.
    Therefore, correspondingly high values of ${\cal W}_\textrm{mm}$ of up to $\sim2.3$ are neglected in the color bar of the panel below to allow the spatial structure to become fully discernible.
  }
  \label{fig:ihsresults}
  ~\vspace*{0.1cm}\\
\end{figure*}

\section{Application of the Model to the IHS}
\label{sec:4}
Taking the plasma beta values obtained with the MHD simulation model (see Section~\ref{sec:mhdmodel}) and using them in the quasilinear model (Section~\ref{qltmodel}) allows us to compute the temperature anisotropy as well as the energy density of the mirror-mode fluctuations throughout the IHS.
Given, however, that the single-fluid MHD model provides a single temperature and thus a single plasma beta value for a given location, further assumptions have to be made.
Therefore, we considered the simulation beta values to correspond to $\beta_\perp$ and computed the quasilinear evolution parametrically for two different \emph{initial} values of the temperature anisotropy $A$.
The latter are motivated as follows.

The first, $A=1.03$, corresponds to the very low ``quasi-isotropic'' values discussed by \citet{Liu-etal-2007}.
Such a low initial value would imply an IHS that is stable with respect to the mirror-mode instability for plasma beta below about 40, i.e.\ for large parts of the computed model heliosphere with the plasma beta values shown in Figure~\ref{fig:heliosphere}.
The actual initial anisotropy, however, might be higher, as is indicated in the simulations by \citet{Liu-etal-2010}.
They show first that, while limited in spatial extent, higher values are possible already for the solar wind protons, and second that the
anisotropy becomes more pronounced when taking into account the presence of pickup protons.
The anisotropy-enhancing effect of the pickup protons has also been emphasized by \citet{Burlaga-Ness-2011} and \citet{Tsurutani-etal-2011a}. 
In order to test this, we checked how large the initial anisotropy must be in order to have mirror-mode instability throughout the heliosheath; we found that $A\gtrsim 1.25$ is needed at $\tau=0$.
In view of the initially (i.e., shortly after ionization and ``pickup'') strongly anisotropic pickup ion distribution \citep[e.g.,][]{Florinski-2009} such values are not at all unreasonable.
The corresponding results are displayed in Figure~\ref{fig:ihsresults}.  
The upper right panel reveals that the quasilinear evolution results in anisotropy values a little above unity across most of the IHS, 
which is consistent with the findings by, e.g., \citet{Liu-etal-2007} and \citet{Fahr-Siewert-2007}.
Also, low anisotropy values correspond to high beta values and vice versa, which appears to be consistent with the marginal stability condition discussed in, e.g., \citet{Hellinger-etal-2006} or \citet{Yoon-2017}.    

The present analysis goes beyond this, however, because within the framework of the outlined quasilinear theory, we can also compute the resulting magnetic energy density of the mirror mode-induced compressive fluctuations and, via Equation~(\ref{deltan}), the associated, locally generated density fluctuations.
The former quantity is shown in the lower left panel of Figure~\ref{fig:ihsresults}.
Clearly, the generation of these fluctuations is significant in large regions of the IHS, particularly also below a latitude of about $45^{\circ}$, i.e., in the region probed by the Voyager spacecraft. 
At high northern latitudes and within the equatorial plane, ${\cal W}_\text{mm}$ is decreasing toward the heliopause, which appears to reflect the distribution of the plasma beta.
The corresponding density fluctuations are shown in the lower right panel.
As a consequence of Equation~(\ref{deltan}), the structure of their distribution follows from that of the magnetic energy density and the temperature anisotropy. 

\begin{figure}[h!]
  \begin{center}
    \includegraphics[width=1.00\columnwidth]{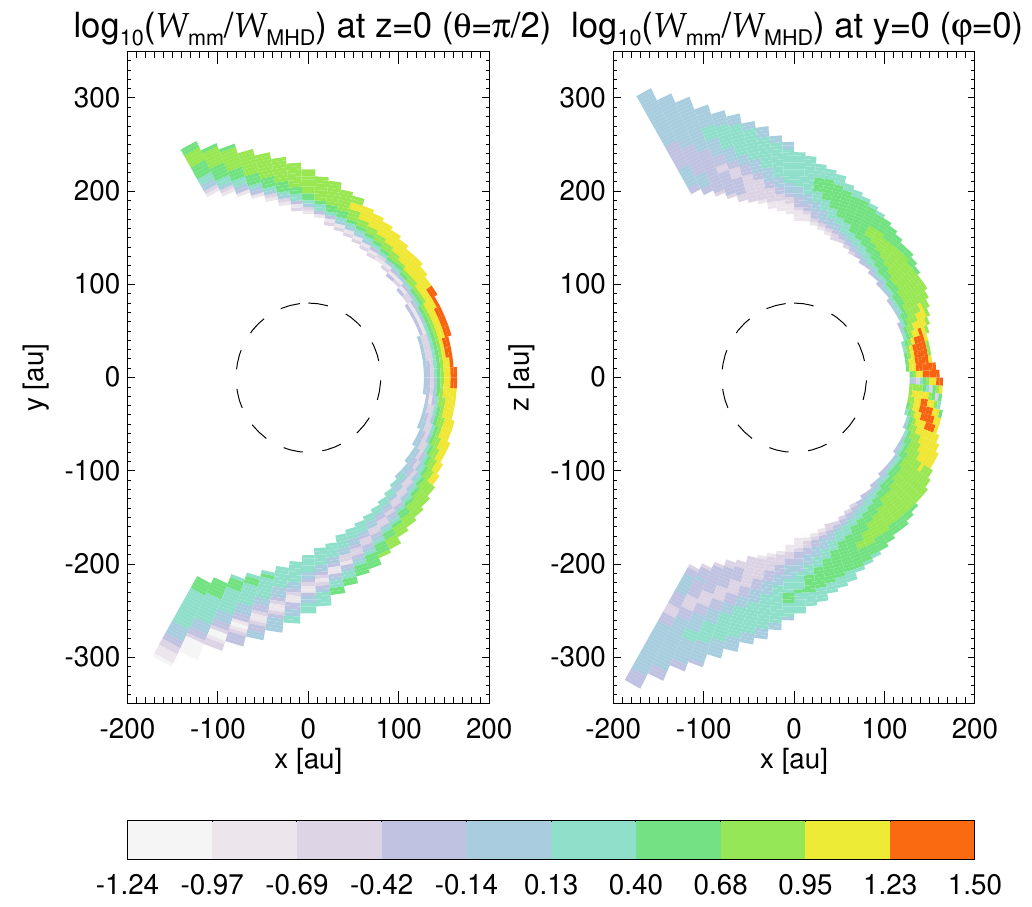}
  \end{center}
  \caption{The ratio of the (normalized) energy density of the compressible magnetic mirror-mode fluctuation ${\cal W}_\text{mm}$ to that of the incompressible MHD fluctuations ${\cal W}_\text{MHD}$ obtained from the simulation described in Section~\ref{sec:mhdmodel}.
    Left panel: equatorial plane. Right panel: meridional plane.}
  \label{fig:enedensratio}
  ~\vspace*{0.2cm}\\
\end{figure}

It is also of interest to compare the magnetic energy density associated with the mirror-mode instability ${\cal W}_\text{mm}$ (see Equation~(\ref{enedens})) with that associated with the MHD fluctuations ${\cal W}_\text{MHD}$, Equation~(\ref{eq:W-mhd}), as obtained from the \textsc{Cronos} simulation described in Section~\ref{sec:mhdmodel} \citep{Wiengarten-etal-2015}. 
For that purpose Figure~\ref{fig:enedensratio} provides the ratio ${\cal W}_\text{mm}/{\cal W}_\text{MHD}$. 
The figure reveals that the compressible fluctuations dominate over much of these cross sections of the IHS, particularly also at the mid-latitudes probed by the Voyager spacecraft.
Only in the equatorial and high-latitude downstream vicinities of the termination shock do the incompressible fluctuations prevail. 

\section{Summary and Conclusions}
\label{sec:5}
With the present study we have quantitatively investigated, for the first time, the generation of compressive fluctuations throughout the inner heliosheath.
It is still unclear whether these fluctuations indirectly observed by the Voyager spacecraft are related to current sheets associated with so-called proton boundary layers \citep{Burlaga-Ness-2011}, to mirror-mode waves resulting from the mirror-mode instability \citep{Liu-etal-2007, Genot-2008}, or even to solitons \citep{Avinash-Zank-2007} or other features. 
\citet{Burlaga-Ness-2011}, however, have offered the hypothesis that observed magnetic ``hole and hump'' structures can be initiated by the mirror-mode instability and subsequently evolve by nonlinear kinetic processes to solitons that form isolated, very slowly propagating, pressure-balanced structures.
Following this idea, we have, on the one hand, tested what level of initial temperature anisotropy, $A:= T_\perp/T_\parallel $, is required in order to have the mirror-mode instability generate compressive fluctuations throughout the whole inner heliosheath.
The determined threshold of $A \gtrsim 1.25$ is reasonable in view of the initially strongly anisotropic velocity distributions of pickup protons that dominate the temperature in this region.
On the other hand, within the framework of quasilinear theory, we were able to compute the energy density of the locally generated corresponding magnetic fluctuations and the associated density fluctuations.
These may serve as source terms in forthcoming models of turbulence transport in the inner heliosheath.

The study may be improved in different ways.
First, the simplifying assumption of bi-Maxwellian distribution functions can be dropped in favor of anisotropic kappa distributions \citep[e.g.,][]{Scherer-etal-2019}.
Second, the model for the global heliosphere that was set up to resemble the main features of the model by \citet{Usmanov-etal-2016} can be improved by considering not only a one-component model of turbulence but a two-component formulation as in, e.g., \citet{Wiengarten-etal-2016}, \citet{AdhikariEA17-NI-2}, \citet{ShiotaEA17}, or \citet{Zank-etal-2017}.
Also, refinements are possible with respect to the plasma structure in the inner heliosheath, such as taking into account the heating of electrons \citep{Engelbrecht-Strauss-2018}, which would change the amount of energy supplied to the protons and probably affect the plasma beta.
Finally, it is desirable to take the compressible mirror-mode fluctuations into account more self-consistently in the MHD modelling.
Such improvements, which are beyond the scope of the present, first quantitative approach, will be the subject of subsequent work.

\newpage
\acknowledgments

We are grateful for support by the German Research Foundation (Deutsche Forschungsgemeinschaft, DFG) through project FI~706/23-1.
P.H.Y. acknowledges NASA Grant NNH18ZDA001N-HSR, NSF Grant 1842643, and Science Award from the GFT Charity, Inc., to the University of Maryland, and the BK21 plus program from the National Research Foundation (NRF), Korea, to Kyung Hee University.

\bibliographystyle{apj}
\bibliography{references}

\end{document}